\documentclass[aps,prd,twocolumn,showpacs,amsmath]{revtex4}
\usepackage[dvips]{color,graphicx}
\usepackage{amsfonts}
\usepackage{amssymb}
\usepackage{latexsym}

\newcommand{\dalm}{\kern1pt\vbox{\hrule height 0.9pt\hbox{\vrule width
0.9pt\hskip 2.5pt\vbox{\vskip 5.5pt}\hskip 3pt\vrule width 0.3pt}\hrule height
0.3pt}\kern1pt}
\newcommand{\ma}[1]{\mbox{$\mathcal{#1}$}}

\newcommand{\lw}[1]{\smash{\lower2.ex\hbox{#1}}}

\begin{document}

%\thispagestyle{empty}

%<<<<<<<<<<<<< TITLE >>>>>>>>>>>>>>>%
\title{On Lovelock vacuum solutions}
%<<<<<<<<<<<<< AUTHOR >>>>>>>>>>>>>>>%
\author{Naresh Dadhich\thanks{Electronic address:nkd@iucaa.ernet.in}}
\email{nkd@iucaa.ernet.in}
%<<<<<<<<<<<<< ADDRESS >>>>>>>>>>>>>>>%
\affiliation{Inter-University Centre for Astronomy \& Astrophysics, Post Bag 4, Pune~411~007, India\\}
\date{\today}

%======================================%
%<<<<<<<<<<<<< ABSTRACT >>>>>>>>>>>>>>>%
%======================================%
\begin{abstract} 
We show that the asymptotic large $r$ limit of all Lovelock vacuum and electrovac solutions with $\Lambda$ is always the Einstein solution in $d \geq 2n+1$ dimensions. It is completely free of the order $n$ of the Lovelock polynomial indicating universal asymptotic behaviour. 
\end{abstract}

%<<<<<<<<<<<<< PACS NUMBER >>>>>>>>>>>>>>>%

\maketitle
Gravity is the universal force which means it links to all that physically exists including the massless particles. That is why it can only be described by the spacetime curvature,  defined by the Riemann curvature tensor. That is what governs its dynamics and it follows in a straightforward manner from the Bianchi differential identity that yields on contraction the divergence free Einstein tensor, $G_{ab}$. It is a second order quasilinear differential operator acting on the metric, $g_{ab}$, an analogue of the usual Laplacian in the Newtonian theory. Without reference to anything else we can straightway write,
\begin{equation}
G_{ab} = \kappa T_{ab} + \Lambda g_{ab}, ~~ T^{ab}_{;b} = 0 
\end{equation}
where the other term on the right is simply a constant relative to the covariant derivative. Now $T_{ab}$ should be the measure of the source which should also be universal for the universal force and hence it can only be the energy momentum distribution. That is indeed shared by all the particles. This is how the Einstein equation for gravitation follows naturally from the Riemann curvature \cite{d1}. Note that $\Lambda$ comes on the same footing as the source, $T_{ab}$ and it is therefore a new constant of nature intimately related to the spacetime structure in some deep and fundamental way. Like $\kappa$, its value has to be determined by observation and it is in fact being measured in the acceleration of the expansion of the Universe \cite{de}! 
\\

Since gravity resides in the curvature of spacetime, nothing should we impose/prescribe on it except some general physical properties like massless free propagation which is in fact required by the universality of gravity - it should reach everywhere. The minimum number of dimensions that could have  massless propagation of gravity is four and that is how we come to our usual four dimensional spacetime. Is there any other physical property that may ask for consideration of dimensions beyond four? There are number of purely classical considerations like embedding into higher dimensional flat spacetime, inherently self interactive nature of gravity and the total charge being zero that do seem to point towards dimension, $d>4$ \cite{d2}. \\

In here, we would refer to high energy behavior of gravity, that would ask for inclusion of higher order terms in the curvature. Usually such corrections are evaluated as perturbations against a fixed background spacetime provided by the low energy solution. Since at high energy, spacetime curvature would also be very strong, hence we cannot resort to the usual perturbative analysis but instead have to consider the situation non-perturbatively. That is, include higher order terms in Riemann curvature, derive the equation of motion and then seek its solution. For the next order, we include square of the Riemann curvature and yet we must have a second order quasilinear (linear in the second derivative) equation for unique evolution with a proper initial data. This uniquely identifies a particular combination, $L_{GB} = R_{abcd}^2 - 4R_{ab}^2 + R^2$, (where $R_{ab}^2 = R_{ab}R^{ab}$) known as the Gauss-Bonnet (GB) Lagrangian. This has remarkable property that the squares of the second derivative get canceled out leaving the equation quasilinear. The variation of this as well as the trace of the Bianchi derivative of a fourth rank tensor which is a homogeneous quadratic in Riemann curvature leads to the analogue of $G_{ab}$, a divergence free $H_{ab}$ \cite{d3}. \\

We define an analogue of $R_{abcd}$ as a homogeneous polynomial \cite{d3} as 
\begin{equation}
\ma F_{abcd} = \ma R_{abcd} - \frac{n-1}{n(d-1)(d-2)} \ma R (g_{ac}g_{bd} - g_{ad}g_{bc}) \label{F}
\end{equation}
where 
\begin{equation}
\ma R_{abcd} = Q_{ab}{}^{mn}R_{mncd}
\end{equation}
and 
\begin{equation}
Q^{ab}{}_{cd} = \delta^{aba_1b_1...a_nb_n}_{cdc_1d_1...c_nd_n}R_{a_1b_1}{}^{c_1d_1}...R_{a_nb_n}{}^{c_nd_n}.
\end{equation}
Here $n$ is the order of polynomial and $Q^{abcd}{}{}{}{}_{;d}=0$. For the quadratic case, $\ma R_{abcd}$ reads as 
\begin{equation}  
\ma R_{abcd} = R_{abmn}R_{cd}{}^{mn} + 4R_[{a}{}^{m}R_{b]mcd} + R R_{abcd} 
\end{equation}
where $\ma R = L_{GB}$. Note that the Bianchi derivative of $\ma F_{abcd}$ does not vanish (that only vanishes for $R_{abcd}$), however its trace does vanish to give 
\begin{equation}
 -\frac{n}{2} \ma F^{cd}{}{}_{[cd;e]} = H_e{}^c_{;c} = 0  
\end{equation} 
where 
\begin{equation}
n(\ma F_{ab} -\frac{1}{2} \ma F g_{ab}) = H_{ab}.
\end{equation}
This is an alternative derivation of $H_{ab}$ which results from the variation of the corresponding $n$-th order term in the Lovelock polynomial which is defined by $Q_{abcd}R^{abcd}$. In the GB quadratic case it takes the form,
\begin{eqnarray}
H_{ab}& =& 2(RR_{ab} - 2R_{a}{}^{m}R_{bm} - 2R^{mn}R_{ambn}\nonumber \\  
&+& R_{a}{}^{mnl}R_{bmnl}) - \frac{1}{2} L_{GB} g_{ab}.
\end{eqnarray}
However $H_{ab}$ is non-zero only for $d>4$ which means GB makes non zero contribution in the equation of motion only in dimension higher than four. This clearly indicates that at high energies gravity cannot remain confined to the four dimensions and the consideration of higher dimensions becomes pertinent and relevant. Its dynamics in higher Dimensions would be governed by $H_{ab} = -\Lambda g_{ab}$. \\

We envision that all matter fields live on four dimensional spacetime, the so called $3$-brane. If matter fields are the gauge vector fields, they would respect the conformal invariance (the fields remain unaffected by the universal change of scale, $g_{ab}  \rightarrow f(x^i)^2 g_{ab}$) only in four  dimensions. All physics except gravity should respect conformal invariance because universal change of scale should not affect all that which does not depend upon the spacetime background. It is therefore reasonable to expect that matter fields remain confined to the usual four dimensional spacetime. In the string theory as well, the matter is supposed to remain confined to the $3$-brane on which the open strings have their endpoints \cite{st}. Only gravity sourced by the high energy effects can penetrate in higher dimensions where it has no matter source. We shall therefore focus on $H_{ab}=-\Lambda g_{ab}$ which describes the high energy gravitational dynamics and is entirely sourced by quadratic terms in the curvatures \cite{note}. We shall therefore seek the solution of this equation for spherically symmetric spacetime. \\ 

Although the Lovelock vacuum solutions are known for long time \cite{whitt} but what we wish to probe here is the universality of the asymptotic large $r$ limit. It turns out that this limit does not much distinguish between the pure Lovelock solution of some order and its Einstein-Lovelock analogue so long as $\Lambda$ is non-zero. In particular, the asymptotic limit of the Einstein-Gauss-Bonnet solution with 
$\Lambda$ has asymptotically the same form as the pure Gauss-Bonnet solution and it is the Einstein solution in $d$ dimensions. However their $r\to 0$ limit is radically  different, for the former the metric remains regular and finite while for the latter it is singular at $r=0$. We would also like to draw attention to an interesting property of spherically symmetric vacuum and electrovac equations that one has ultimately to solve a single first order equation not withstanding the enhanced nonlinearity of the Lovelock gravity. \\  

We shall begin with the GB vacuum equation, 
\begin{equation}
H_{ab} = -\Lambda g_{ab}
\end{equation} 
for the spherically symmetric metric, 
\begin{equation}
ds^2= e^\nu dt^2 - e^\lambda dr^2 - r^2d\Omega_{d-2}^2
\end{equation}
where $d\Omega_{d-2}^2$ is the metric on a unit $(d-2)$-sphere. In general $\nu, \lambda$ are functions of both $t$ and $r$, however as shown in \cite{whitt1} the $t$ dependence drops out as usual and it then suffices to take them as functions of $r$ alone. To begin with we have $H^t_t = H^r_r$ that immediately determines $\nu = -\lambda$. With this, let us write the non-zero components of $H^a_b$ for the above metric and they read as follows: 
\begin{equation} 
H^t_t = H^r_r = -\frac{d-2}{2r^4}(1-e^{-\lambda}) \bigl(2re^{-\lambda}\lambda^{\prime} + (d-5)(1-e^{-\lambda}) \bigr)  
=-\Lambda 
\end{equation}
\begin{eqnarray}
H^\theta_\theta&=& \frac{1}{2r^4} \Bigl[r^2e^{-\lambda}(1-e^{-\lambda})(-2\lambda^{\prime \prime} + \lambda^{\prime^2}) 
- r^2e^{-2\lambda}(3-e^{\lambda})\lambda^{\prime^2} \nonumber \\
&+& (d-5)(1-e^{-\lambda})\bigl(-4re^{-\lambda}\lambda^{\prime} - (d-6)(1-e^{-\lambda}) \bigr) \Bigr] \nonumber \\
&=& -\Lambda 
\end{eqnarray}
where a prime denotes derivative w.r.t $r$ and all the angular components are equal. \\

First let us note that the above two equations are not independent and it can easily be seen that the latter is a derivative of the former which was first shown for the usual four dimensional gravity in \cite{d-sch}. It would therefore suffice to integrate the former alone to get the general solution. This is what it should be because there is only one function, $\lambda$, to be determined. Eq (11) could be written as 
\begin{equation}
(r^{d-5}f^2)^\prime = \frac{2\Lambda}{d-2}r^{d-2}
\end{equation}
which readily integrates to give 
\begin{equation} 
e^{-\lambda} = F = 1-f, ~~ f^2 = \frac{k}{r^{d-5}} + \Lambda_1 r^4
\end{equation}
where $2\Lambda/{(d-1)(d-2)} = \Lambda_1$. This is the general solution of the pure Gauss-Bonnet vacuum which has been obtained by solving the single first order equation. \\ 

Let us take the large $r$ limit of this solution,
\begin{equation} 
F =1-\sqrt{\Lambda_1}r^2-\frac{K}{r^{d-3}} 
\end{equation}
where $K = k/{2\sqrt{\Lambda_1}}$. This is the Schwarzschild-dS solution for a $d$-dimensional spacetime. On the other hand let us look at the Einstein-Gauss-Bonnet solution (which is obtained by summing over $n=1,2$ in Eq (13)) \cite{bd}, 
\begin{equation} 
F = 1+\frac{r^2}{2\alpha}[1-\sqrt{1+4\alpha(\frac{M}{r^{d-1}}+\Lambda)}] 
\end{equation}
that would also approximate for large $r$ to 
\begin{equation} 
F = 1 - \Lambda r^2 - \frac{M}{r^{d-3}}.  
\end{equation}
Thus the two solutions perfectly agree in the large $r$ limit. It should however be noted that for the former the presence of $\Lambda$ is essential for this limit to exist. \\ 
 
Now we go to the general case and we write $G_{ab}^{(n)}$ for the differential operator resulting from the $n$-th term in the Lovelock polynomial and in particular, $G_{ab}^{(1)}$ is the Einstein tensor and $G_{ab}^{(2)} = H_{ab}$ of the Gauss-Bonnet. Note that for the spherically symmetric vacuum as well as for electrovacuum, null energy condition, $G_{ab}^{(n)}k^ak^b = 0, ~ k^ak_a = 0$, will always hold good and thereby implying $\nu = -\lambda$. Hence there would be left only one parameter to be determined. As above, we would again have the analogue of Eq (12) as derivative of the analogue of Eq (13) which would now read as 
\begin{equation}
(r^{d-2n-1}f^n)^\prime = \frac{2\Lambda}{d-2}r^{d-2} + \frac{2q^2}{r^{2(d-2)}}
\end{equation}
and it would readily integrate to give 
\begin{equation}
f^n = \Lambda_1 r^{2n} + \frac{k}{r^{d-2n-1}} - \frac{Q^2}{r^{2(d-n-2)}} 
\end{equation}
where $Q^2 = 2q^2/{(d-2)(d-3)}$. Here we have included the Maxwell charge on the particle and this is the general electrovac solution for any $n = 1, 2, ...,n$ in the Lovelock polynomial. For $Q=0$, it agrees with the known solution \cite{whitt1}. \\ 

Let us take the large $r$ limit of the above solution and it would read as
\begin{equation} 
F =1-\Lambda_1^{1/n} r^2-\frac{K}{r^{d-3}} + \frac{e^2}{r^{2(d-3)}}
\end{equation}
where now $K=k/{n\Lambda_1^{1/n}}$ and $e^2=Q^2/{n\Lambda_1^{1-1/n}}$. This is a charged black hole in $d$-dimensional dS/AdS (for even $n$ only dS while for odd $n$ it could be dS/AdS with the sign of $\Lambda$). The corresponding Einstein-Lovelock solution is simply obtained by summing over $n$ in Eq (13) with a coupling coefficient for each $n$. It is however expected that asymptotically the solution should tend to the Einstein solution in $d$ dimensions. We have seen that above for $n=2$ and we have also verified it for $n=3$ \cite{dehghani}. \\

Thus we establish that asymptotically the pure Lovelock for a given order $n$ and the Einstein-Lovelock ($\sum_n G_{ab}^n$) tend to the same limit of the Einstein solution in $d$-dimensional dS/AdS spacetime. It is understandable that the higher order Lovelock contributions should wean out asymptotically, however what is rather interesting and intriguing is the fact that even the order $n$ in the Lovelock polynomial does not matter so long as $\Lambda$ is  included. That is, the large $r$ limit is free of $n$ indicating a universal asymptotic behaviour. \\  

The higher order Lovelock terms are supposed to account for the high energy corrections which would be dominant in the 
$r\to0$ limit. In this limit, the pure Lovelock and Einstein-Lovelock solutions indeed have radically different behaviour. For the former, the metric is singular while for the latter it is regular and finite everywhere. This is the distinguishing Lovelock corrections effect. However the curvature diverges but by one power less compared to the pure Lovelock solution. The high energy Lovelock corrections thus tend to weaken/smoothen the singularity. On the other hand, at the low energy large $r$ end, it is always the Einstein gravity that overrides and is therefore universal. \\
 
The other point we would like to emphasize is the remarkable property that spherically symmetric vacuum and electrovac equation ultimately reduces to a single first order equation (Eq (18) with sum over $n$) whose integration is trivial. In the general Lovelock case with all the orders included and each order contributing a coupling coefficient, the difficult part is to solve the  algebraic polynomial in $f$. It becomes highly involved and combersome. Further it is argued that it becomes very difficult to extract meaningful physical information in the context of black hole thermodynamics. To circumvent this difficulty, the method of dimensional continuation has been employed to study higher order black holes \cite{zan}. It prescribes a  relation between the different Lovelock coefficients and that leads to the solution in the form, $F = 1 + r^2/l^2 - (K/r^{d-2n-1})^{1/n}$ which is quite different from the solutions we have considered above. Here the AdS part and the gravitational potential, which is purely due to the pure $n$-th order, have been separated and it does not have the Einstein limit (which would require potential to go as $K/r^{d-3}$) for a $d$-dimensional black hole is AdS spacetime.  \\

For the familiar Einstein gravity, we have $n=1$ and the solution is then the Schwarzschild-deSitter for $d=4$ and the BTZ black hole for $d=3$ ~\cite{btz}. It should be noted that $G_{ab}^{(n)}$ is non-zero only in dimension, $d\geq 2n+1$. In $d = 2n+1$ and $\Lambda=0$, the potential, $f$ in the solution turns constant which represents a deficit solid angle for $d>3$ and it is deficit angle for $d=3$. Even when both $g_{tt}$ and $g_{rr}$ are constant yet spacetime is non-flat because the former can be absorbed by redfinition of $t$ while the latter cannot be without it appearing in $r^2\Omega_{d-2}^2$. Though the angle deficit (as is the case for $d=3$) does not produce non-zero curvature but the solid angle deficit (defecit angles also occur in codimension 2 braneworld gravity \cite{left}) does and hence the spacetime is non-flat. \\

In particular for the Gauss-Bonnet quadratic case in $5$ dimension, the constant potential will generate the Einstein stresses, $G^t_t=G^r_r = -3k/r^2, G^\theta_\theta = -k/r^2$. Its analogue in $4$-spacetime has $G^t_t=G^r_r = k/r^2, G^\theta_\theta = 0$ that match with the asymptotic limit of the global monopole stresses and the spacetime has been interpreted as describing a global monopole \cite{bv}. It turns out that these stresses also agree with the asymptotic limit of $5$-dimensional global monopole ~\cite{d-ghosh}.  That is the pure Gauss-Bonnet vacuum solution in $5$-spacetime is effectively equivalent to a $5$-dimensional  global monopole. It is easy to see that the $n$-th order Lovelock will similarly produce a global monopole in $(2n+1)$-dimensional spacetime. Further, the stresses for global monopole has the universal character that they will always fall as $1/r^2$ because they are produced by the deficit solid angle.  \\

In all these considearions the most important question is, why is gravity supposed to propagate in higher dimensions when all other fields remain confined to the four dimensions? How is gravity different from the others? The fundamental difference is in its universality - its linkage to all particles and that is why it cannot be removed globally. The answer to all its peculiarities has to stem from this remarkable and unique feature. It is this that requires that it could only be described by the spacetime curvature. That immediately implies that it cannot obey the conformal invariance which all other gauge vector fields do because for gravity, the metric is its potential. Hence the scaling $g_{ab} \rightarrow f^2 g_{ab}$ for gravity is not innocent change of universal scale but is a non-trivial change in its dynamics. If the conformal invariance is to be adhered to, the vector fields can exist only in four dimensions. On the other hand, gravity with massless free propagation can exist in any dimension $\geq4$. Another general feature of the  classical fields is that the total charge must vanish. For gravity, energy-momentum is the charge which is unipolar and  always positive. The question is, how to counter it, but counter it must to have total charge zero? The only possibility is the field it produces must have charge of opposite polarity. That is why the gravitational interaction energy must be negative and the field always attractive. But the negative charge (field energy) is spread all over the space and is not localizable. If we integrate over the whole space, it would perfectly balance the masspoint \cite{adm}. However in the neighbourhood of a masspoint, there would be overdominance of positive charge and hence the field must propagate in higher dimension but with diminishing field strength. Because as it propagates in higher dimension, its past light cone will enclose more and more of negative charge of the field and so its field strength in higher dimension will go on diminishing. Thus garvity may propagate in higher  dimension but not deep enough \cite{d2}. That is, for the matter living in $4$-spacetime where gravity it produces has massless free propagationin while the propagation in higher diemsnions is however with the diminishing strength. This is a crucial new feature of gravitational dynamics in higher dimensions. The picture that emerges is similar to that of the Randall-Sundrum braneworld gravity \cite{rs} where zero mass propagation remains confined on the brane and the bulk has massive propagation. This is a very intuitive and enlightening classical argument but it has not yet been formulated in a rigorous quantitative manner. \\   

For propagation of gravity in higher dimensions, it is interesting to draw a parallel with the dynamics of confinement of the strong force where the opposite happens for the field strength. It becomes stronger as the field propagates out to keep it confined \cite{das}. Elsewhere I had suggested that there was a kind of complimetarity between gravity and the strong force \cite{uni}. The former has universal linkage and universal reach everywhere while the latter has neither. In some appropriate way, there should perhaps exist a duality relation between them. The well known AdS/CFT correspondence \cite{mal} seems to augur and resonate well with this kind of suggestion. \\ 
       
It therefore appears strongly suggestive that gravity does penetrate in higher diemsnion and if its dynamics there is described by the GB (next order in Lovelock) vacuum equation, it effectively generates a constant potential or a global monopole in the next higher dimensional spacetime. If we stick to the paradigm of unique evolution of dynamics from a given initial data which requires the quasilinearity of the equation, gravitational dynamics in higher dimensions has to be described by $G_{ab}^{(n)}$ resulting from the Lovelock polynomial. The overall picture that emerges is as follows: if matter remains confined to the $3-$brane and that is where gravity has the usual Einstein dynamics while its dynamics in the next higher dimension is  governed by the GB vacuum equation and it generates only a  constant potential, which can be interpreted as a global monopole, in the five dimensional de Sitter spacetime. We have focused on five dimensions and the quadratic GB polynomial simply  because this is the next higher dimension to the $3$-brane on which the matter resides. The important question now is to find the effect of the constant potential(global monopole)-dS bulk on the $3$-brane gravity. \\

We should however emphasize that it is the purely classical consideration of high energy effects that points to higher dimension and thereby to the Gauss-Bonnet (in general Lovelock) gravity. Any quantum theory of gravity must therefore include the high energy limit of classical gravity. That is, it must first approximate to the Gauss-Bonnet and then subsquently to the Einstein gravity. This seems to suggest that the Gauss-Bonnet gravity may be the intermediatory limit to quantum gravity \cite{d2}. In particular it should include the pure Gauss-Bonnet vacuum solution obtained here; i.e. a five dimensional de Sitter spacetime with solid angle deficit or constant potential. This is a definitive suggestion for a candidate quantum theory of gravity. \\  

Finally at low energy end, all Lovelock solutions in any form always approach the Einstein solution thereby establishing the universality of the asymptotic behaviour of vacuum and electrovac spacetime. 

Most humbly I dedicate this work to the fond memory of Professor P. C. Vaidya who had always been a source of inspiration for me.

%{\it Acknowledgment:}

%======================================%
%<<<<<<<<<<<<< REFERENCES >>>>>>>>>>>>>%
%======================================%


\begin{thebibliography}{9}
\bibitem{note}
The Einstein-GB vacuum equation would however read as $G_{ab} + \alpha_{GB} H_{ab} = -\Lambda g_{ab}$ where $\alpha_{GB}$ is the GB coupling constant. In absence of maatter in higher dimensions, it would perhaps be appropriate to consider only  the GB vacuum part $H_{ab} = -\Lambda g_{ab}$.
\end{thebibliography}
\end{document}